\documentstyle[aps,floats,epsfig]{revtex} 
\twocolumn

\newcommand{\dt}{\Delta \tau}

\newcommand{\tJ}{$t$-$J$\ }

\begin{document}
\wideabs{
\title{Thermodynamics of the $t$-$J$ Ladder:
A Stable Finite Temperature Density Matrix Renormalization Group 
Calculation}
\author{Beat Ammon $^{1,2,3}$\cite{ABA},
Matthias Troyer $^{2,3}$\cite{AMT}
T.M. Rice $^2$ and N. Shibata$^{4}$}
\address{$1$ SCSC, Eidgen\"ossische Technische Hochschule, CH-8092 Z\"urich,
        Switzerland \\
	$2$ Theoretische Physik, 
        Eidgen\"ossische Technische Hochschule, CH-8093 Z\"urich,
        Switzerland \\
	$^3$ ISSP, University of Tokyo, 7-22-1 Roppongi, Minato-ku, Tokyo 106,
	 Japan \\
	 $^{4}$Institute of Applied Physics, University of Tsukuba, 
	Ibaraki 305-3006, Japan }
\maketitle
\begin{abstract}
Accurate numerical simulations of a doped $t$-$J$ model
on a two-leg ladder are presented for the particle number,
chemical potential, magnetic susceptibility and entropy in the
limit of large exchange coupling on the rung using a finite 
temperature density matrix renormalization group (TDMRG) method. This
required an improved algorithm to achieve numerical stability down
to low temperatures. The thermal dissociation of hole pairs and of the 
rung singlets are separately observed and the evolution of the hole
pair binding energy and magnon spin gap with hole doping is determined.
\end{abstract}
\pacs{71.10.Fd,71.27.+a,74.25.Bt,02.60.Cb}
} 
Standard quantum Monte Carlo methods for the simulation of fermions
are limited to relatively high temperatures due to the fermion sign
problem. The density matrix renormalization group method (DMRG)
\cite{DMRG} allows simulations of large clusters but is limited to
groundstate properties. In this Letter we report on a finite
temperature DMRG (TDMRG) \cite{TDMRG,FTDMRG} method using an improved
numerically stable algorithm to simulate a strongly interacting
fermion system down to low temperatures.

The TDMRG method applies the DMRG to the quantum transfer matrix (QTM)
in the real space direction \cite{TM}. 
In the TDMRG  iterations the QTM is enlarged in the imaginary time direction 
and iterates to lower temperatures at fixed Trotter time steps 
$\Delta\tau$. This is in contrast to the
DMRG method in which the system grows in the real space direction.
The TDMRG has the advantage that the free energy and other
thermodynamic quantities for the {\it infinite} system can be obtained 
directly from the largest eigenvalue 
and the QTM and of the corresponding eigenvector. 

The system we examine is a two-leg $t$-$J$ ladder model in the
limit where the exchange interaction across the rungs ($J'$) is
large compared to the value along the legs($J$) and to the isotropic
hopping integral $t$. The ground state properties of this model
at low hole doping have been analyzed previously by exact 
diagonalization of small clusters \cite{PRBTroyer,ladderreview}. In this limit
$J'\gg J,t$ the thermal dissociation of hole pairs and the excitation of 
triplet magnons can be distinguished. These
strong coupling processes are a good test for any method.

In this Letter we present accurate results for the magnetic 
susceptibility $\chi$, the particle number $n$ and the entropy density $s$
in the grand canonical ensemble as a function of chemical potential 
$\mu$ and temperature $T$ and then remap $\chi(\mu,T)\rightarrow
\chi(n,T)$ to obtain the $T$-dependence at constant density.

Previous versions of the TDMRG method for fermions \cite{FTDMRG} have
suffered from numerical instabilities due to the non-Hermiticity of
the QTM and the corresponding density matrices which are constructed
from the right and left eigenvector of the largest eigenvalue of the
QTM.  These numerical instabilities grow as the number of states kept
is increased or the filling is changed away from half-filled bands. We
have identified the loss of biorthonormality between the left and
right eigenvectors $(v_i^{(l)},v_j^{(r)})=\delta_{{ij}}$ of the
density matrix as the source of the problem. The biorthogonal but
normalized eigenvectors $v_i^{(l)}/||v_i^{(l)}||_2$ and
$v_j^{(r)}/||v_j^{(r)}||_2$ have to be multiplied with a factor
$\left[(v_i^{(l)},v_i^{(r)})/(||v_i^{(l)}||_2
||v_i^{(r)}||_2)\right]^{-1/2}$, to become {\it biorthonormal}, which
leads to severe loss of precision due to roundoff errors if the
overlap between these vectors is small. These near-breakdowns occur
especially often in conjunction with the second numerical problem,
spurious small imaginary parts of (nearly) degenerate eigenvalue
pairs.  This latter problem can be solved by using the real and
imaginary components of the corresponding complex conjugate
eigenvector pairs and discarding the imaginary part of the
eigenvalues, which are artifacts of roundoff errors and are only of
the order of the machine precision.  To circumvent the loss of
precision in the former problem of nearly orthogonal eigenvectors our
algorithm uses an iterative re-biorthogonalization step
\cite{ReBiOrtho} for the eigenvectors kept, which stabilizes the
method for all temperatures.  Technical details of the algorithms will
be presented elsewhere
\cite{StableFiniteT}. 
\begin{table}
\begin{center}
  \caption[*]{Free energy density $f$ for noninteracting spin $S=1/2$
  fermions on a chain with $\mu=0$ at a temperature $T=0.1$ after a
  hundred DMRG steps ($\dt=0.1$), where energies are given in units of
  the hopping integral $t$.  The first column (I) is the original
  TDMRG algorithm \cite{TDMRG,FTDMRG}, and the second column (II) is
  our improved method including the re-biorthogonalization step.
  Cases where the algorithm becomes numerically unstable are denoted
  with $\dagger n$, where $n$ is the number of DMRG steps that could
  be performed successfully.}\label{TabBreakdown}
\begin{tabular}{l|c c c}
algorithm & I & II \\
\hline$m=10$ & -0.65500& -0.65500 \\
$m=20$ & $\dagger$ 16 &  -0.67225 \\
$m=30$ & $\dagger$ 39 &  -0.67413 \\
$m=40$ & $\dagger$ 9 &   -0.67441 \\
$m=50$ & $\dagger$ 7 & -0.67454 \\
$m=60$ & $\dagger$ 50 &  -0.67457 \\
$m=80$ & $\dagger$ 8 &   -0.67462 
\end{tabular}
\end{center}
\end{table}
In Tab.~\ref{TabBreakdown} we show results of numerical stability 
tests of the original \cite{TDMRG,FTDMRG} and our improved algorithm
for the case of noninteracting spin $S=1/2$ fermions in one dimension.
For this simple fermionic model the original TDMRG method
becomes numerically unstable whenever more than about $m=10$
states are kept, thus severely restricting the achievable accuracy.
The improved algorithm presented here, on the other hand, is 
always numerically stable and achieves much higher accuracy.
The test example clearly demonstrates the need for numerical stabilization
in the simulation of fermionic models.  

The results of the stabilized TDMRG method are accurate and
unbiased, with errors only originating from the finite size of the
Trotter time steps and the truncation in the DMRG
algorithm. The latter are usually very small if the
number of states kept, $m$, is large enough, and the former
can be eliminated by extrapolating
$\dt\rightarrow 0$ by fitting to a polynomial in $\dt^2$. We have
used Trotter time steps from $\dt t=0.01$ to $\dt t=0.2$, and
$m$ between $m=40$ and  $m=60$. 
We made use of spin conservation symmetry, the
subspace of zero winding number and the reflection symmetry of the
ladder along the rungs to optimize the calculations
and reduce numerical errors.

Thermodynamic quantities such as the internal energy $U$, the hole density 
$n_{h}(=1-n)$ 
and the magnetic susceptibility $\chi$ have been determined directly from
the eigenvectors of the transfer matrix \cite{TM,troyer_tm}. This is
preferable to taking numerical derivatives of the free energy density 
obtained from the largest eigenvalue of the QTM.

The low temperature properties of a doped $t$-$J$ two-leg ladder in the
limit $J'\gg J,t$ are determined solely by the singlet hole pairs 
(HP) \cite{PRBTroyer}. They form a hard core boson gas with 
a bandwidth of $4t^{*}$, with $t^{*}={2t^2}(J'-4{t}^2/J')^{{-1}}$ in 
second order perturbation theory. Neglecting a weak nearest neighbor 
attraction, the HP fluid can be mapped to an ideal Fermi gas 
in this one dimensional geometry. As the temperature $T$ is increased 
the HPs dissociate into two quasiparticles (QP), each consisting of
a single electron with spin $S=1/2$ in a rung bonding state. Each QP 
propagates with a bandwidth of $2t$ so that in the limit of low hole 
doping $n_{h}\ll 1$ the HP binding energy is $E_{B}=J'-4t+4t^{2}
(J'-4{t}^2/J')^{{-1}}$. The gas of QPs with density $n_{QP}(T)$ 
contributes to the spin susceptibility as a nondegenerate gas of 
$S=1/2$ fermions. A second contribution comes from the thermal 
excitation of singlet rungs to a triplet magnon state. The activation 
energy for a magnon $\Delta_{M}$ ($=J'-J+J^{2}/2J'$ in second order 
perturbation theory in the limit $n_{h}\rightarrow0$ 
\cite{gap_hb})
is larger than that of the QPs ($\Delta_{QP}=E_{B}/2$) but since the 
density of QPs is limited by the hole density ($n_{QP}(T)\le n_{h}$), 
the temperature evolution of $\chi(n,T)$ is determined largely by the magnons 
at low doping.

\begin{figure}
   \begin{center}
   \epsfxsize=\linewidth
   \epsffile{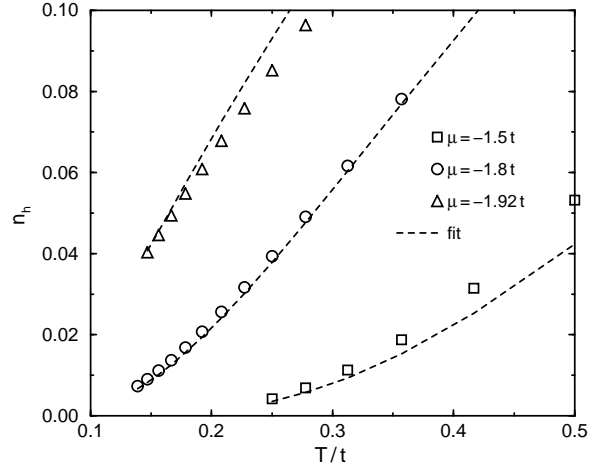}
   \end{center} 
   \caption[*]{Hole density $n_h$ as a function of electron chemical potential
   $\mu$ and temperature $T$ in the strong coupling regime $J=t/2=J'/10$.
   The dashed lines are fits to a hard core boson model for the hole 
   pairs. Note, the HP chemical potential is $-2\mu$.}
\label{fig:nhole}
\end{figure}
We now turn to the presentation of our finite 
temperature results obtained using the
improved TDMRG algorithm for $J=t/2=J'/10$ and compare them to expectations 
based on the above discussion of this strong coupling regime.
As the calculations were performed in the grand canonical ensemble
we first present results for the hole density $n_{h}(\mu,T)$. A selection
of our results is presented in Fig.~\ref{fig:nhole} including a fit
to a hard core boson model for the HPs:
\begin{equation}
\epsilon_{HP}^{k} = \epsilon_{HP}+2t^{*}\cos k +2\mu n_{h}.
\end{equation}
Fitting the
data for $n_{h}<0.1$ at temperatures $T<0.5t$ we obtain an estimate for
the center of the band for HPs at $\epsilon_{HP}=4.82(6) t$ and a 
bandwidth $4t^{*}=1.5(2)t$. The minimum energy to add a HP to an 
undoped ladder is $\epsilon_{HP}-2t^{*}= 4.1(1) t$, in good agreement 
with values from the finite clusters ($\epsilon_{HP}\approx 4.71t$,
$4t^{*}\approx 1.494t$ \cite{PRBTroyer}).

\begin{figure}[ht]
   \begin{center}
   \epsfxsize=\linewidth
   \epsffile{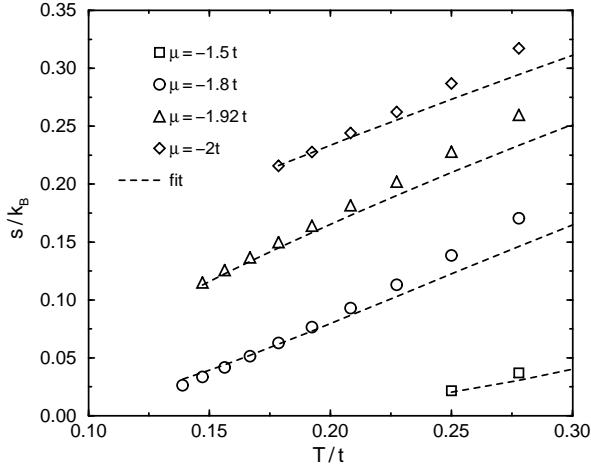}
   \end{center} 
   \caption[*]{Entropy density $s$ as a function of chemical potential
   $\mu$ and temperature $T$ in the strong coupling regime $J=t/2=J'/10$.
   The dashed lines are the values for the same hard core boson model 
   as in Fig.~\protect{\ref{fig:nhole}}, using the parameters obtained
   in that fit.}
\label{fig:entropy}
\end{figure}
A further confirmation of the validity of this hard core boson model
for the HPs comes from considering the low temperature 
entropy density per site $s$, determined
from the free energy density $f$ and the energy density $u$ as 
$s=(u-f)/T$. As can be seen in Fig.~\ref{fig:entropy} the entropy at 
$T<0.3t$ and low doping ($n_{h}<0.1$) is also well described by the hard 
core boson model for the HPs. 

\begin{figure}[hbt]
   \begin{center}
   \epsfxsize=\linewidth
   \epsffile{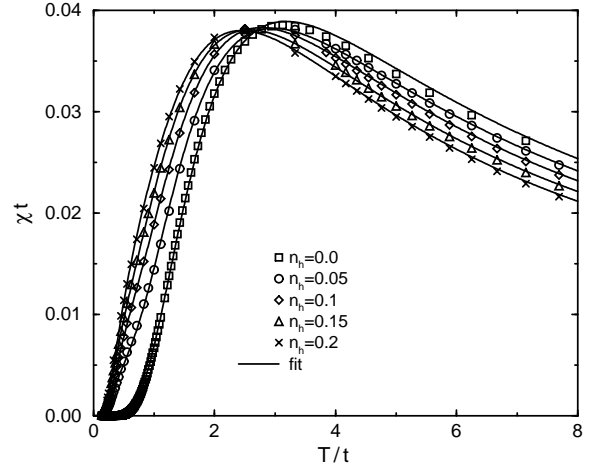}
   \end{center} 
   \caption[*]{Uniform magnetic
   susceptibility per site $\chi t$ of the \tJ ladder for $J=t/2=J'/10$ and
   different hole-densities $n_h$. The symbols denote the
results of the TDMRG algorithm, and the solid
lines are the fitted curves according to Eq.~(\ref{LadderFit}). 
The fitting parameters are listed in Tab.~\ref{TabChiLadder}.}
\label{SuscLad}
\end{figure}
At higher temperatures the thermal 
dissociation of HPs into two independent QPs and the thermal 
excitation of magnons from rung singlets govern the thermodynamics. 
These processes show up in the spin susceptibility 
$\chi(T)$, which is easiest to interpret in the canonical ensemble with fixed 
hole  density $n_{h}$. Therefore we use $n_{h}(\mu,T)$ to
remap $\chi(\mu,T)\rightarrow\chi(n_{h},T)$. The values of $\chi(\mu,T)$ 
were calculated by measuring the magnetization $\langle 
S^{z}(T)\rangle$ in the presence of a small external field 
$h/t=5\times 10^{-3}$. The results for $\chi(n_{h},T)$ 
appear in Fig.~\ref{SuscLad}.

At high temperatures $T\gg J'$, $\chi$ follows 
a Curie-law for
free spins $\chi=(1-n_{h})/4T$, and it decreases
when the temperature is lowered below the magnon-gap $\Delta_M
\approx 4.13t$.  The maximum of the peak is
shifted towards lower $T$ with increasing doping, 
indicating a reduction of the magnon gap due to interactions with 
holes. Simultaneously the magnon bandwidth is enhanced, indicating 
that the energy of a localized magnon is not much changed by the holes.
At very low temperatures of $T < 0.5t$ we can see a
second exponential decrease of $\chi$ with a  smaller gap, which we
attribute to the recombination of QPs into HPs at temperatures below 
the QP-gap, $\Delta_{QP}=E_{B}/2$. Note the magnitude of this
contribution increases with $n_{h}$.

A quantitative description of $\chi(n_{h},T)$ can be given
by adding separately the contributions of the QPs $\chi_{QP}$ 
and of the magnons $\chi_{M}$, i.e.:
\begin{equation}\label{LadderFit}
\chi(n_{h},T)=\chi_{QP}(n_{h},T) +  \chi_M(n_{h},T).
\end{equation}
$\chi_{QP}$ is approximated by the value for free spins 
$\chi_{QP}=n_{QP}(T)/4T$ with a temperature
dependent density of the QPs determined by the energy dispersion of the QPs 
$\varepsilon_{QP}^{k}=\Delta_{QP}+a_{QP}(1+\cos k)/2$ with $\beta=1/T$:
\begin{equation}\label{NQP}
n_{QP}=\frac{n_h}{\pi}\int_{-\pi}^{\pi} \!dk\, 
\frac{1}{e^{\beta \varepsilon_{QP}^{k}}+1}.
\end{equation}

The density of rungs occupied by two spins at low temperatures where
all holes are bound in HPs is $1-n_{h}$ but exciting QPs reduces
the number of such rungs by one for each QP so that the rung density 
is then $1-n_{h}-n_{QP}$. Our approach to a model for
$\chi_{M}$ is simply to scale the form for undoped ladders proposed by 
Troyer {\it et al.} \cite{troyer_tm} by this two-spin rung density leading to
\begin{equation}\label{TroyerFit}
\chi_M=(1-n_{h}-n_{QP})\beta\frac{z(\beta)}{1+3z(\beta)},
\end{equation}
where $z(\beta)=
\int_{-\pi}^{\pi}\!dk (2\pi)^{{-1}}\exp(-\beta\varepsilon^k_M)$, and
$\varepsilon^k_M=[\Delta_M^2+4a_M(1+\cos k)]^{1/2}$ \cite{gopalan}. 

\begin{table}
\begin{center}
  \caption[*]{Gap of the spin $S=1/2$ quasi-particles $\Delta_{QP}$ and
        magnon gap $\Delta_M$, as well as the parameters $a_{QP}$ ($a_M$) 
	which determine the bandwidth of the quasi-particles (magnons)
	obtained by fitting Eq.~(\ref{LadderFit}) to our TDMRG data
        for different hole densities $n_h$.}
\begin{tabular}{l|c c c c}
$n_h$ & $\Delta_{QP}$ & $\Delta_{M}$ & $a_{QP}$ & $a_M$ \\
\hline
0.0     &   -   & 4.1(1) &  -   & 0.7(1) \\
0.025   & 0.7(1) & 3.4(1) &  0.9(2) & 1.6(2) \\
0.05    & 0.8(1) & 3.3(1) &  0.9(2) & 1.8(2) \\
0.1     & 1.0(1) & 3.3(1) &  0.6(3) & 1.7(2) \\
0.15    & 0.9(1) & 3.2(1) &  1.3(2) & 2.0(2) \\
0.2     & 0.9(1) & 3.2(1) &  1.4(2) & 2.0(2) \\
\end{tabular}\label{TabChiLadder}
\end{center}
\end{table}

The parameters obtained by a fit of this model to the TDMRG data are
shown in Tab.~\ref{TabChiLadder}.  The main change upon doping is the
decrease of the magnon gap $\Delta_M$
\cite{gap_hb,PoilblancScalapino,moregap} due to interactions between
the magnons and QPs.  Due to hybridization with higher lying bands the
QP bandwidth $a_{QP}$ is also reduced from the leading order
perturbation result $a_{QP}=2t$, but the QP gap $\Delta_{QP}$ is in
reasonable agreement with the second order perturbative estimate of
$0.98t$.  The increase of $\Delta_{QP}$ (or equivalently the binding
energy $E_B$) with $n_{h}$, can be attributed to an effective
repulsion between the QPs and HPs.  A similar increase of the QP gap
$\Delta_{QP}$ was found in Ref.~\cite{PoilblancScalapino}.  This is an
issue which warrants further investigations.


\begin{figure}
   \begin{center}
   \epsfxsize=\linewidth
   \epsffile{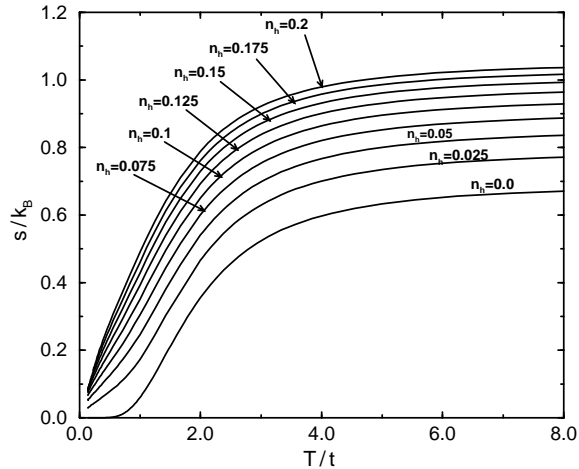}
   \end{center} 
\caption[*]{   Entropy density of the doped \tJ ladder with
   $J=t/2=t'/2=J'/10$ and different hole-densities $n_h$. 
}
\label{EntLadder}
\end{figure}
Finally, in Fig.~\ref{EntLadder} we show the entropy density $s$, 
remapped in the same way to the canonical ensemble of
fixed $n_{h}$. In the limit of $T\rightarrow \infty$ 
$s_\infty=(1-n_h)\ln 2 -n_h \ln n_h-(1-n_h)\ln(1-n_h)$.
At $T/t=20$ the entropy has acquired between $99.4\%$ of its maximal
value $s_\infty$ for $n_h=0.025$ and $99.7\%$ for $n_h=0.2$.  Below
the magnon gap $\Delta_M$, the entropy decreases exponentially, for
the undoped Heisenberg ladder down to $s=0$. In the presence of hole
doping, the exponential decrease shows a crossover to a linear
decrease at low temperatures, as is expected for Luther-Emery
liquids. This behavior is consistent with the hard core boson model
proposed for the magnons. Quantitative fits are however better 
performed on $s(\mu,T)$, as we did earlier, due to added 
uncertainties arising from the remapping to constant hole doping.

In conclusion, we have developed the first numerically stable
TDMRG algorithm for fermionic systems. This has
enabled us to calculate accurate results for the
magnetic susceptibility $\chi$ and the entropy density $s$
of the doped \tJ ladder with strong exchange on the rungs down to low
temperatures in the thermodynamic limit of infinite system size.
The system we have studied has two crossovers as the spins
bind in singlet pairs and the holes in hole pairs. These crossovers
can be clearly seen in the numerical data and demonstrate
that this form of the TDMRG can be successfully used to
reliably simulate strongly interacting fermions over a wide
temperature range.

Finally, very recently Rommer and Eggert report TDMRG calculations
for a spin chain with impurities \cite{rommer} which uses the same method
as we do to overcome the problems caused by complex eigenvalues, but 
they did not introduce the re-biorthogonalization which for fermionic 
models, we find is essential.

We wish to thank E. Heeb, M. Imada, and X. Wang for useful discussions.
Most of the calculations have been performed on the DEC 8400 5/300's of the
C4-cluster at ETH Z\"urich.

\end{document}